# The $X$ boson ($M_X \sim 10^{14}$ GeV) mediation of the HI relativistic reactions generated by RHIC and LHC

M. Kozlowski

Physics Department, Warsaw University, Warsaw, Poland

**Abstract**

In this paper the model for the p-p accretion process in HI reactions is described. The mediation by $X$ boson ($M_X \sim 10^{14}$ GeV) is outlined. The Heaviside equation for the thermal processes in HI scattering is obtained and solved. The possibility of the detection of the collective excitation in HI generated by RHIC and LHC

1. Introduction

The interaction of the relativistic heavy ion is full of enigma. Strictly speaking we even do not know what sort of medium is created at first step of the reaction: fluid or gas [1].

In this paper starting from the low energy nuclear reactions we propose the Heaviside equation for the thermal processes in HI interactions. We calculate the relaxation time and speed of the thermal disturbance. The solution of the Heaviside equation shed new light on the initial step of the HI processes.

2. The model equation

When an energetic projectile (e.g. heavy ion interacts with the target nucleus) hot nuclear drops (HND) are formed. In paper[2] the process of the cooling of HND through the accretion of cold nucleons was proposed.

Assuming the simple model for the expansion of a hot nuclear drop

$$\frac{dA}{dt} = \gamma \qquad (1)$$

(A is the mass number of the hot drop and $\gamma$ is the constant) the cooling curve for hot nuclear matter can be calculated [2]. Comparison of the experimental data with the accretion model calculation yields the following value for the constant: $\gamma = 0.05$ N/($10^{-15}$ m/c) = $1.5 \times 10^{22}$ s$^{-1}$. Assuming for the hot nuclear drop the radius – mass ratio as for a cold nucleus, $R = r_o A^{1/3}$, $r_o = 1.2 \times 10^{-15}$ m, one obtains from Eq.(1)

$$\frac{dR}{dt} = \frac{r_o^3}{3R^2} = H_N(R) \times R, \qquad H_N = \frac{\gamma}{3A}. \qquad (2)$$

Formula (2) was written in the form of the "Hubble law" with the "Hubble constant" $H_N$. For a contemporary nuclear "Universe" $A$, the mass number, can be taken as $A \leq 300$. With $A \sim 300$ (the limit for superheavy elements), one obtain for $H_N$

$$H_N = 1.68 \times 10^{19} \text{ s}^{-1}. \qquad (3)$$

It is interesting to compare the value of $H_N$ to the value of the cosmological (gravitational) Hubble constant at present epoch, $H_o \sim 75$ km s$^{-1}$ Mpc$^{-1}$ = $2.43 \times 10^{-18}$ s$^{-1}$:

$$H_N / H_o = 6.9 \times 10^{36}. \qquad (4)$$

The very large dimensionless number obtained can be compared to the ratio of electric force between two protons to the gravitational force between them. To that aim let us denote by $F_{pp}^C$ the Coulomb force and by $F_{pp}^N$ the Newtonian force. We will use the SI system of units. In the SI system

$$F_{pp}^C = \frac{q_e^2}{4\pi\varepsilon_o} \frac{1}{r^2}, \qquad F_{pp}^N = \frac{Gm_N^2}{r^2}, \qquad (5)$$

where $q_e$ = elementary electric charge = $1.6 \times 10^{-19}$ C, $q_e^2 / 4\pi\varepsilon_0 = 2.31 \times 10^{-28}$ J m, $G = 6.67 \times 10^{-11}$ m$^3$ kg$^{-1}$ s$^{-2}$ and $m_N = 1.66 \times 10^{-27}$ kg.

Considering formula (5) we obtain

$$\frac{F_{pp}^C}{F_{pp}^N} = \frac{q_e^2}{4\pi\varepsilon_o Gm_N^2} = 1.24 \times 10^{36}. \qquad (6)$$

We conclude that

$$\frac{q_e^2}{4\pi\varepsilon_o Gm_N^2} \sim \frac{H_N}{H_o} \qquad (7)$$

that is, both large dimensionless numbers are of the same order.

In the following considering the Dirac Large Number Hypothesis [3] we put:

$$\frac{H_N}{H_o} = \frac{q_e^2}{4\pi\varepsilon_o G_N m_N^2}. \qquad (8)$$

With the definition of the age of the Universe, $t_o = H_o^{-1}$ formula (8) gives

$$H_N = \frac{q_e^2}{4\pi\varepsilon_o G_N m_N^2} t_o^{-1}. \qquad (9)$$

With the analogy to cosmological time $t_o$ we define characteristic time $t_N = H_N^{-1}$ as

$$t_N = \frac{4\pi\varepsilon_o G_N m_N^2}{q_e^2} t_o = 3.3 \times 10^{-19} \text{ s}. \qquad (10)$$

The time $t_N$ can be considered as the lifetime of HND, that is, time elapsed from its "creation". The maximum velocity of expansion of the HND surface is equal to $c$, the speed of light. When the expansion velocity reaches $c$, the growth of the hot drop ceases. The limit value of the HND radius is equal to

$$R_N = ct_N \sim 1 \text{ Å} = 0.1 \text{ nm}. \qquad (11)$$

The HND with radius $R = 0.1$ nm contains $\sim 10^{14}$ nucleons and has mass $M_{HND} = (a_o/r_o)^3 m_p = 10^{14}$ GeV/$c^2$.

It is interesting to observe that the calculated mass of the HND is the same as the mass of the massive vector boson $X$ which mediates proton decay [4].

It seems quite reasonable to argue that the same $X$ boson mediates the accretion of the nucleons in relativistic heavy ion reactions, Fig. 1.

The amplitude for the accretion can be written as

$$A(q^2) \sim \frac{1}{q^2 + 4M_X^2} \qquad (12)$$

The range for accretion "force" is equal $\Delta l$, where

$$q\Delta l \sim \hbar \qquad (13)$$

From formula (13) we calculate $\Delta l$

$$\Delta l \sim 10^{-15} \text{ fm} \qquad (14)$$

for $M_X \sim 10^{14}$ GeV.

The characteristic time, the relaxation time for proton – proton scattering via boson $X$ is

$$\tau \sim \Delta t \sim \frac{10^{-15} \text{ fm}}{c} \sim 10^{-38} \text{ s}. \qquad (15)$$

Having in hand the relaxation time $\tau$ and speed of interaction propagation $c$ we can formulate the thermal energy transport equation [5]

$$\frac{1}{c^2}\frac{\partial^2 T}{\partial t^2} + \frac{1}{D}\frac{\partial T}{\partial t} = \nabla^2 T, \tag{16}$$

where $D$ is the diffusion coefficient $D = \hbar/M_X$.

Equation (16) is the Heaviside equation for thermal processes mediated by $M_X$.

3. The solution of the Heaviside equation

The Cauchy initial condition for Eq.(16) can be written as

$$T(x,0) = 0 \qquad T(0,t) = f(x). \tag{17}$$

For initial condition (17) the solution of Eq.(16) has the form

$$T(x,t) = \left\{ f(t - \tfrac{x}{c})\exp[-(\rho x/c)] + \frac{\sigma x}{c}\int_{x/c}^{t} f(t-y)\exp[-y\rho]\frac{I_1[\sigma(y^2 - (x^2/c^2))^{1/2}]}{(y^2 - (x^2/c^2))^{1/2}} dy \right\}$$
$$\times H(t - \tfrac{x}{c}).$$

(18)

In formula (18) $\rho = \sigma = (1/2\tau)$ and $H(t - x/c)$ is the Heaviside function.

As can be seen from formula (18) for very short time we can observe the propagation specific thermal wave with velocity $c$. This open new field of investigation for HI interaction, generated by RHIC and LHC

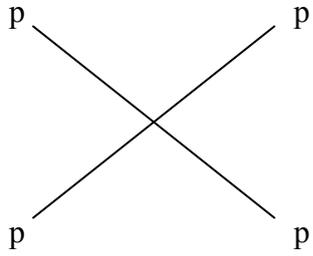
(a)

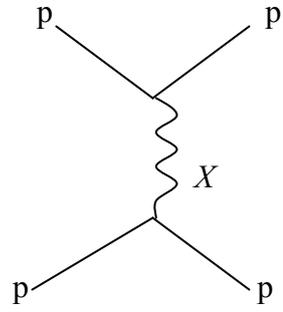
(b)

Fig. 1. (a) Zero range p – p scattering.
(b) Finite range ($\Delta r \sim 10^{-15}$ fm) p – p scattering diagram.